\documentclass{article}
\usepackage{dcase2024_techrep,amsmath,graphicx,url,times,booktabs, tabularx}
\usepackage{soul,color}
\usepackage{caption,setspace,float}
\usepackage{subcaption}

\def\etal{\emph{et al}.\ }

\title{Leveraging Reverberation and Visual Depth Cues for Sound Event Localization and Detection with Distance Estimation}

\twoauthors
  {Davide Berghi}
    {CVSSP, University of Surrey,\\
     Guildford, UK \\
     davide.berghi@surrey.ac.uk}
  {Philip J. B. Jackson}
    {CVSSP, University of Surrey,\\
     Guildford, UK \\
     p.jackson@surrey.ac.uk}

\begin{document}

\ninept
\maketitle

\begin{sloppy}

\begin{abstract}
This report describes our systems submitted for the DCASE2024 Task 3 challenge: Audio and Audiovisual Sound Event Localization and Detection with Source Distance Estimation (Track B).
Our main model is based on the audio-visual (AV) Conformer, which processes video and audio embeddings extracted with ResNet50 and with an audio encoder pre-trained on SELD, respectively.
This model outperformed the audio-visual baseline of the development set of the STARSS23 dataset by a wide margin, halving its DOAE and improving the F1 by more than 3x. Our second system performs a temporal ensemble from the outputs of the AV-Conformer. We then extended the model with features for distance estimation, such as direct and reverberant signal components extracted from the omnidirectional audio channel, and depth maps extracted from the video frames. While the new system improved the RDE of our previous model by about 3 percentage points, it achieved a lower F1 score. This may be caused by sound classes that rarely appear in the training set and that the more complex system does not detect, as analysis can determine. To overcome this problem, our fourth and final system consists of an ensemble strategy combining the predictions of the other three. Many opportunities to refine the system and training strategy can be tested in future ablation experiments, and likely achieve incremental performance gains for this audio-visual task.

\end{abstract}


\section{Introduction}
\label{sec:intro}

Sound Event Localization and Detection (SELD) consists of simultaneously detecting and classifying the active sound sources over time (sound event detection (SED)) while predicting their position or direction of arrival DOA \cite{Adavanne:2019:SELDnet}.
This task is essential for a variety of applications, such as human-robot interaction, augmented reality, navigation, smart home, and security, to name a few. 
Over time, researchers have been addressing more and more challenges related to the task, including the detection of polyphonic sounds \cite{Adavanne2019_DCASE}, simultaneous same-class events and moving sources \cite{politis:2020:DCASE}, and external interfering sounds that must not be detected \cite{politis:2021:DCASE}.
This year, the challenge also involves predicting the distance of the active sources \cite{krause2024SELDdist}.
While SELD has been typically formulated as an audio-only task, a parallel audio-visual track has been included in the past two editions of the DCASE challenge, leveraging the Sony-TAu Realistic Spatial Soundscapes 2023 (STARSS23) dataset \cite{Shimada2023STARSS23AA}.
STARSS23 includes 360° video recordings spatially and temporally aligned to the acoustic sound-field captured by the microphone array. This allows the exploration of SELD as a multimodal audio-visual problem (AV-SELD). The two modalities are complementary and can be beneficial to the task: vision provides high spatial accuracy whereas audio can detect occluded objects.

This report describes our systems submitted to the audio-visual track of the challenge. Specifically, we adopted two main models and methods. The first extends our previous work on SELD where an audio-visual (AV) conformer takes as input the concatenation of audio and video embeddings extracted with pre-trained encoders \cite{Berghi:2024:ICASSP24}. The extension consists of adapting the existing model to support distance estimation, as described in Section~\ref{sec:AVconf}.
In Section~\ref{sec:depthModel}, the second method investigates additional input features and extends the model architecture to improve distance estimation and take more advantage of the visual modality.
We report our results on the STARSS23 development set \cite{Shimada2023STARSS23AA} in Section~\ref{sec:experiments}, including a temporal ensemble and an ensemble of all three variants, improving substantially over the challenge baselines.
Section~\ref{sec:conclusion} concludes.

%To solve the problem of concurrent same-class events, an event independent network (EINv2) was proposed \cite{cao:2021:EINv2}. It presents a track-wise output format where each sound event is assigned to an output track. To avoid the track permutation problem, EINv2 is trained with Permutation Invariant Training (PIT). Inspired by this, Shimada \textit{et al.} \cite{Shimada:2022:multiACCDOA} proposed multi ACCDAO (m-ACCDOA) vectors as SELD output representation. Since an ACCDOA vector encodes the activity of the sound event in the length of the vector, the authors proposed class-wise ADPIT (ADPIT), enabling track permutations and duplicating the ACCDOA vectors for training. 

%In terms of audio architecture for SELD, many recent works adopted the conformer network \cite{Gulati2020ConformerCT}. The conformer was initially proposed for speech recognition but lately, it was adopted to achieve SOTA results in SELD too \cite{Wang:2023:dcase23}.
%In the SELD literature, it is often preceded by a convolutional backbone to first extract high-level representations of the input audio features, for instance, Wang \etal \cite{Wang:2023:ACS} employed a ResNet-Conformer.

\section{Audio-Visual Conformer}
\label{sec:AVconf}

As depicted in the left part of Fig.\,\ref{fig:models}, this method extracts audio and visual feature embeddings with an audio and a visual encoder, respectively. The embeddings are then concatenated and processed by a Conformer module with 4 layers. Finally, the output features are fed to two fully connected layers to predict multi-ACCDDOA vectors \cite{krause2024SELDdist}. Such representation is the extension of the multi-ACCDOA vectors proposed by Shimada \etal \cite{Shimada:2022:multiACCDOA}, adopted to include distance predictions. 
Specifically, at each time frame, the output presents $N=3$ tracks, each predicting 4 positional values (x, y, z, and distance), for each of the $C=13$ classes, resulting in a 156-dimensional vector. 
We employed the hyperbolic tangent as the activation function for the x, y, z predictions, while ReLU for the distance.
In this report, we will refer to this system as per ``\textit{AV-Conformer}''.
The same audio-visual architecture and methods were presented in our recent work \cite{Berghi:2024:ICASSP24}. The main difference with the system presented for this challenge consists in the integration of the distance estimation and the use of the new metrics. 
Below, we describe the audio and visual encoders, and the data augmentation and pre-training strategies.

\begin{figure*}[tb]
\centerline{\includegraphics[width=\textwidth]{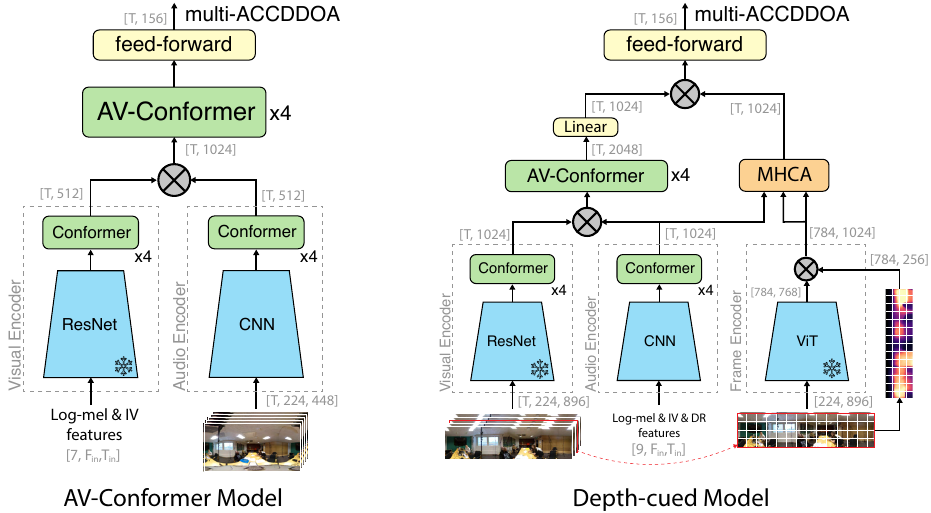}}
\caption{Diagram of the two main architectures submitted to the challenge. On the left, the AV-Conformer with the respective audio and visual encoders. On the right, the Depth-cued model that includes a frame encoder employed to extract visual features from the central frame. The Depth-cued model leverages cubemap views whereas the AV-Conformer equirectangular views. The snowflake symbol indicates that the weights of ResNet50 and ViT are fixed during training.}
\label{fig:models}
\vspace{-4mm}
\end{figure*}

\subsection{Audio Encoder}
\label{sec:audioEnc}

The audio encoder takes as input acoustic features extracted from the FOA spatial sound. We employed intensity vectors (IV) in the log-mel space concatenated with the log-mel spectrograms extracted from the FOA channels, yielding 7-dimensional input features with shape $7\times T_{in}\times F_{in}$, where $T_{in}$ corresponds to the number of temporal bins and $F_{in}$ the frequency bins.
The audio encoder includes a CNN architecture followed by a Conformer \cite{Gulati2020ConformerCT}. The CNN presents 4 convolutional blocks with residual connections. Each block consists of two 3$\times$3 convolutional layers followed by average pooling, batch normalization, and ReLu activation. The average pooling layer is applied with a stride of 2, halving the temporal and frequency dimension at each block. The resulting tensor of shape $512\times T_{in}/16\times F_{in}/16$ is then reshaped and frequency average pooling is applied to achieve a $T_{in}/16\times 512$ dimensional feature embedding. $T_{in}$ is chosen so that $T_{in}/16$ matches the label frame rate (10 labels per second).
The feature embedding is further processed by a Conformer module with 4 layers and 8 heads each. The size of the kernel for the depthwise convolutions is set to 51.

\subsection{Visual Encoder}
\label{sec:visualEnc}

As visual encoder, we employed a ResNet-Conformer. 
Specifically, we fed each video frame to ResNet50 \cite{He:2016:resnet} at a frame rate of 10 fps. In such a way, we extract a number of frame embeddings that match the label frame rate as well as the audio embedding temporal resolution. We then process the frame embeddings with a Conformer module identical to the one employed in the audio encoder.
The video segments used as inputs to the visual encoders were resized to 448$\times$224p. The original ResNet50 model is pre-trained on squared 224$\times$224p frames and its last layer is a global average pooling applied on a 7$\times$7 feature map. Since, in our case, the input horizontal dimension is 448, the resulting feature map has shape 14$\times$7. Therefore, we replaced the last global average pooling with a 7$\times$7 average pooling kernel with stride equal to 7, obtaining two output vectors that are then concatenated. 
Before being fed to the Conformer module, the output of ResNet50 is reduced to 512 dimensions with a linear layer.

\begin{figure}[tb]
\begin{minipage}{0.49\columnwidth}
\includegraphics[width=\columnwidth]{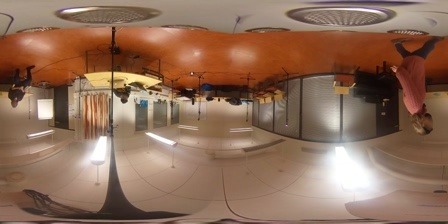}
\vspace{-7mm}
\captionof*{figure}{\footnotesize$\hat{\phi}=\phi-\dfrac{\pi}{2}, \hat{\theta}=-\theta$}
\end{minipage}\hfill
\begin{minipage}{0.49\columnwidth}
\includegraphics[width=\columnwidth]{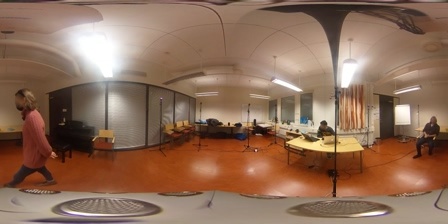} 
\vspace{-7mm}
\captionof*{figure}{\footnotesize$\hat{\phi}=-\phi-\dfrac{\pi}{2}, \hat{\theta}=\theta$}   
\end{minipage}

\begin{minipage}{0.49\columnwidth}
\includegraphics[width=\columnwidth]{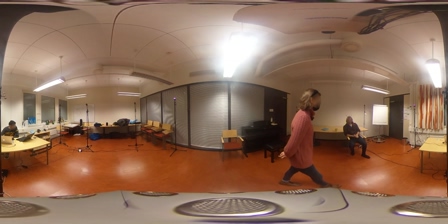}
\vspace{-7mm}
\captionof*{figure}{\footnotesize$\hat{\phi}=\phi, \hat{\theta}=\theta$}
\end{minipage}\hfill
\begin{minipage}{0.49\columnwidth}
\includegraphics[width=\columnwidth]{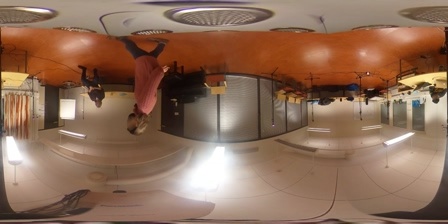} 
\vspace{-7mm}
\captionof*{figure}{\footnotesize$\hat{\phi}=-\phi, \hat{\theta}=-\theta$}   
\end{minipage}

\begin{minipage}{0.49\columnwidth}
\includegraphics[width=\columnwidth]{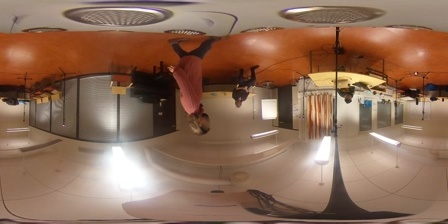}
\vspace{-7mm}
\captionof*{figure}{\footnotesize$\hat{\phi}=\phi+\dfrac{\pi}{2}, \hat{\theta}=-\theta$}
\end{minipage}\hfill
\begin{minipage}{0.49\columnwidth}
\includegraphics[width=\columnwidth]{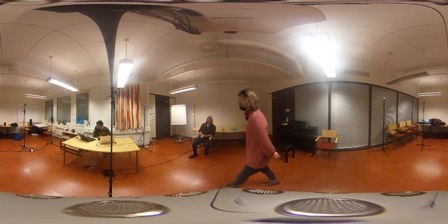} 
\vspace{-7mm}
\captionof*{figure}{\footnotesize$\hat{\phi}=-\phi+\dfrac{\pi}{2}, \hat{\theta}=\theta$}   
\end{minipage}

\begin{minipage}{0.49\columnwidth}
\includegraphics[width=\columnwidth]{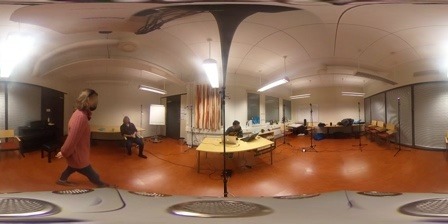}
\vspace{-7mm}
\captionof*{figure}{\footnotesize$\hat{\phi}=\phi+\pi, \hat{\theta}=\theta$}
\end{minipage}\hfill
\begin{minipage}{0.49\columnwidth}
\includegraphics[width=\columnwidth]{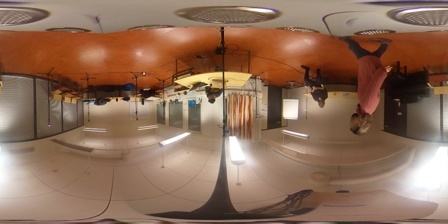} 
\vspace{-7mm}
\captionof*{figure}{\footnotesize$\hat{\phi}=-\phi+\pi, \hat{\theta}=-\theta$}   
\end{minipage}
\captionof{figure}{Examples of visual transformation in relation to the respective DOA augmentation for ``fold3\_room6\_mix006.mp4''.}
\label{fig:AVCS}
\vspace{-4mm}
\end{figure}

\subsection{Data Augmentation and Pre-Training}
\label{sec:augmentation}
The audio CNN-Conformer was pre-trained on SELD employing the simulated data generator script provided for the DCASE 2022 challenge \cite{Politis2022DCASE}. This allowed the generation of $\sim$30h of spatial recordings, including noiseless and noisy versions. The ResNet50 model employed to process 2D frames is available with the Torchvision library and it is pre-trained on ImageNet \cite{ImageNet:2009:dataset}.
The synthetic SELD dataset used to pre-train the audio encoder was augmented by a factor of 8 with the audio channels swap (ACS) technique \cite{Wang:2023:ACS}. 
For the AV-SELD dataset, we also augmented the visual modality consistently with the ACS transformation, i.e., the audio-visual channel swap (AVCS) \cite{Berghi:2024:ICASSP24,Roman:2024:avseld}. This generates new video frames by flipping and rotating the original ones, creating an effective audio-visual spatial transformation. An example of a transformed frame is shown in Fig.\,\ref{fig:AVCS}

\section{Depth-Cued Model}
\label{sec:depthModel}

The other main system submitted to the challenge differs significantly from the previous one in many aspects, including model architecture and size, audio and visual data pre-processing, and audio pre-training. Consequently, their overall performances cannot be directly compared. A step-by-step ablation study would be necessary to understand the impact of each change.
We refer to this method as per ``\textit{Depth-cued model}'', since particular attention was given to incorporating audio and visual depth features.

\subsection{Direct and Reverberant Audio Features}
\label{sec:derRevFeat}

While log-mel spectrograms and intensity vectors (IV) are effective input features for the SED and DOAE subtasks, they are not ideal for tackling the distance estimation task. 
Distance cues can be extracted from the relationship between the direct and reverberant components of the captured audio signals. Specifically, the later tail of the reverberant signal (late reverberation) carries information about the sound's apparent distance.
We decided to include both direct and reverberant components (``DR'', for compactness) extracted from the Omnidirectional audio channel to the set of input features, in the form of two additional log-mel spectrograms. %For compactness, we refer to these features as per ``DR''.
In order to estimate the direct sound, we applied the Weighted Prediction Error (WPE) dereverberation algorithm \cite{Takuya:2012:WPE}. Specifically, we adopted the python implementation of the WPE algorithm released by Drude \etal \cite{Drude:2018:nara_wpe} (taps=60; delay=5; iterations=5).
The reverberant component is then estimated by taking the difference between the original and the direct signal in the temporal domain. 
RD features are then concatenated to the 4 log-mel spectrograms and the 3 intensity vectors used before, producing audio input features with shape $9\times T_{in}\times F_{in}$.

\subsection{Cubemap Conversion and Depth Map Extraction}
\label{sec:cubemap}

The 360° videos of STARSS23 are in the form of 2:1 equirectangular views. 
Such representations produce heavy distortions that become more severe closer to the borders of the frame. We argue that this might penalize the ability of the visual encoder to understand the scene since ResNet50 is pre-trained on frontal views.
To this end, we converted the video frames to cubemap representations. Such image representation consists of mapping the 360° image data onto the six faces of a cube.
Since the sound events are primarily located within the four horizontal faces (left-right-front-back) and little information can be obtained from the top (ceiling) and the bottom (EigenMike) faces, we kept only the horizontal faces. The resulting frames present an aspect ratio of 4:1, with resolution 896$\times$224p.
Note, the conversion to cubemap is applied after the AVCS technique is performed to the equirectangular views.

To leverage the visual modality in support of the distance estimation task, we extracted depth features from video frames too. Specifically, we applied the recent ``Depth Anything'' model \cite{depthanything} to generate depth views of the scene.
Fig.\,\ref{fig:depth} shows (a) a frame with the original equirectangular representation, (b) its cubemap representation, and (c) the depth map extracted from the horizontal faces of the cubemap representation.

\subsection{Depth-cued Model's Architecture}
\label{sec:architecture}

Part of the Depth-cued model architecture presents the same audio and visual encoders and Conformer unit employed in the AV-Conformer model. However, the input frames used with the Depth-cued model are cubemap views, and the audio encoder is pre-trained on the synthetic mixtures provided by the challenge's organizers and takes as input also DR features. Additionally, both audio and visual encoders produce 1024-dimensional embeddings.

An additional frame encoder is included too. It consists of a vision transformer (ViT) \cite{Dosovitskiy2020ViT} pre-trained on ImageNet \cite{ImageNet:2009:dataset}.
The input tokens fed to the ViT are obtained by dividing the central frame into patches of 16$\times$16p. Therefore, a total of $14\times56=784$ tokens are extracted from the 896$\times$224p frame, and each token presents $16\times16\times3=768$ dimensions, where 3 corresponds to the number of color channels.
The same patch extraction technique is applied to the depth map frame feature too, which only presents a single channel. The depth patches are then concatenated to the output of the ViT to produce 782$\times$1024 dimensional features.
These are employed to generate a set of Key-Value pairs for a multi-head cross-attention (MHCA) unit applied against the audio features (Queries).

The main motivation for employing the ViT and the MHCA unit is to try to leverage the spatial accuracy provided by the visual modality. Unlike ResNet50 which downsamples the spatial resolution of the input frames with the risk of losing spatial information, we argue that the ViT should preserve such information encoded within the processed patches. 
The outputs generated by the AV-Conformer and by the MHCA are then concatenated and fed to a feed-forward network that predicts the output multi-ACCDDOA vectors \cite{krause2024SELDdist}.

\section{Experiments}
\label{sec:experiments}

\begin{figure}[tb]
\begin{minipage}{\columnwidth}
\includegraphics[width=\columnwidth]{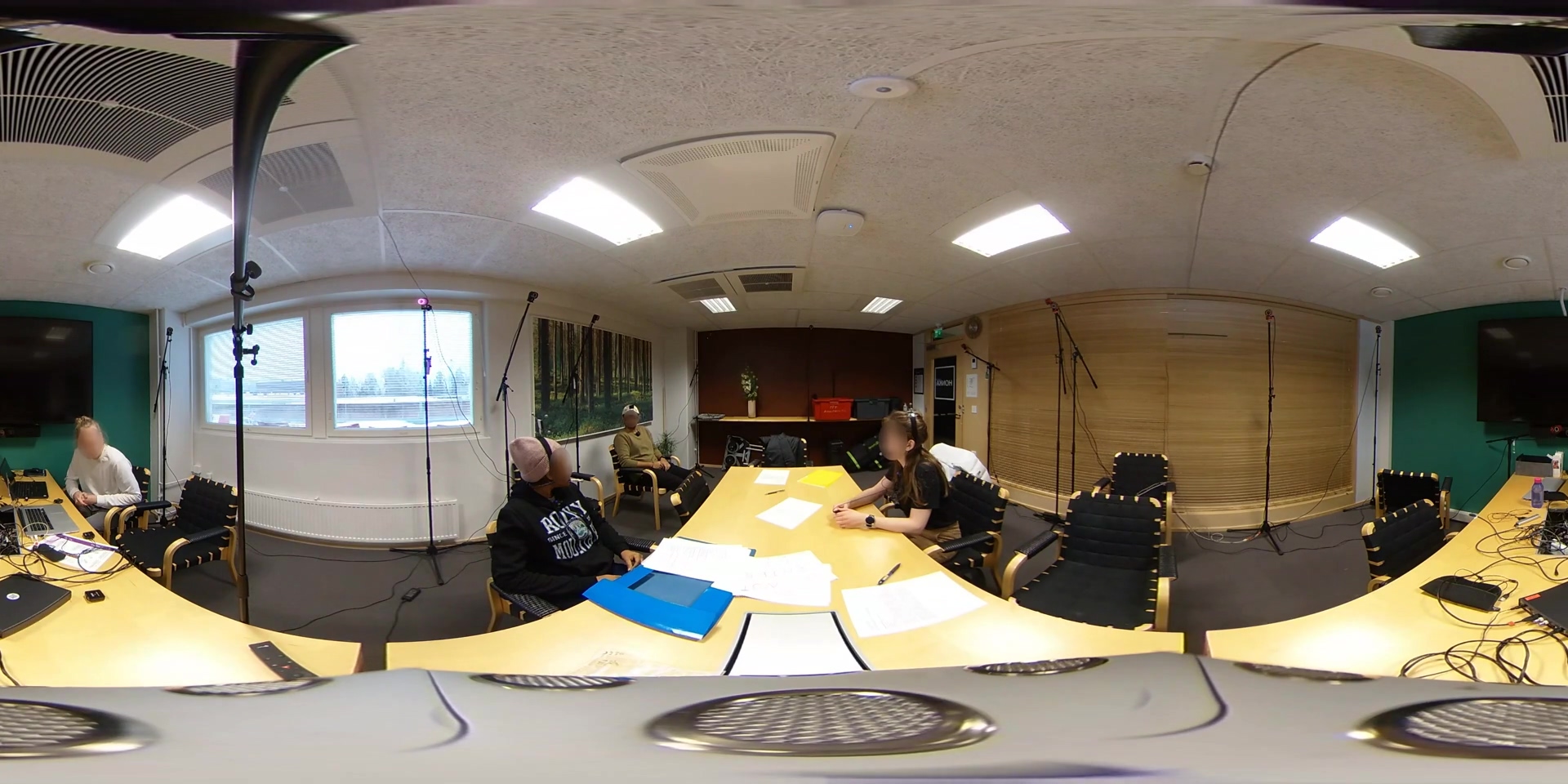}
\vspace{-7mm}
\captionof*{figure}{\footnotesize (a) Original equirectangular view}
\end{minipage}\hfill
\begin{minipage}{\columnwidth}
\includegraphics[width=\columnwidth]{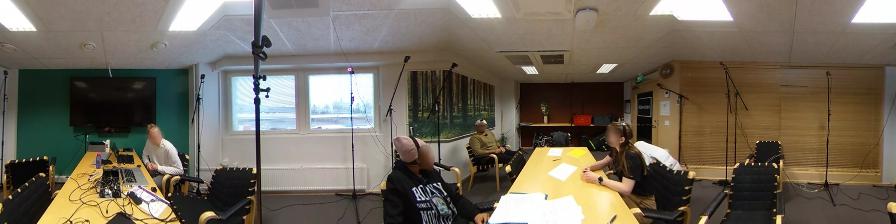} 
\vspace{-7mm}
\captionof*{figure}{\footnotesize (b) Horizontal faces of the cubemap view}   
\end{minipage}

\begin{minipage}{\columnwidth}
\includegraphics[width=\columnwidth]{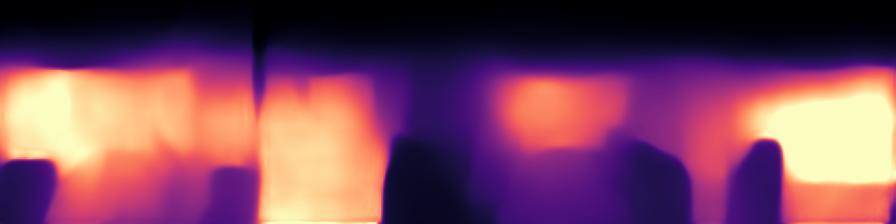}
\vspace{-7mm}
\captionof*{figure}{\footnotesize (c) Depth map features}
\end{minipage}\hfill

\captionof{figure}{Examples of cubemap transformation and depth map features for ``fold3\_room13\_mix003.mp4''. Note how by removing the top and bottom faces from the cubemap representation the EigenMike and a good portion of the ceiling are no longer in the frame.}
\label{fig:depth}
\vspace{-3mm}
\end{figure}

\subsection{Implementation Details}

To train our models, we divided the dataset into chunks of 3 seconds, extracted at steps of 1s for training and with no overlap for testing.
An STFT with 512-point Hann window and hop size of 150 samples generates spectrograms discretizing the 3-second audio chunks (24kHz) into 480 temporal bins ($T_{in}$). We used 128 frequency bins to generate log-mel spectrograms and intensity vectors (IV) in log-mel space.   
We trained our models with batches of 32 inputs and Adam optimizer for 40 epochs, then we selected the best epoch. The learning rate is set to 0.00005 for the first 30 epochs, then and it is decreased by 5\% every epoch.
The metrics adopted for the evaluation are the ones proposed in the DCASE 2024 Task3 Challenge \cite{Shimada2023STARSS23AA}.

\subsection{Ensemble strategies}

Inference is conducted on 3-second video segments without any overlap, meaning each segment starts exactly 3 seconds after the previous one. To enhance spatial accuracy, we tested a temporal ensemble (TE) strategy. For this, inference was performed with a 1-second hop size, generating 3 predictions for each second of the sequence (except for the first 2 seconds). A sound event is considered active only if at least 2 out of the 3 predictions detect it.
To determine if the sounds detected by different predictions are from the same event, we applied a spatial threshold of 15°. Predictions are considered related if their positions are within this threshold. The average of their x, y, z, and distance predictions is employed as the final temporal ensemble.
This approach improved spatial accuracy but reduced the F1 score. As a result, we submitted a single system using TE, specifically applied to the AV-Conformer model.

As a result, our submitted systems are (1) the AV-Conformer method, (2) the AV-Conformer with TE, and (3) the Depth-cued model. 
We found that the Depth-cued model rarely detects ``Water tap'' and ``Bell'' sounds, and it never detects ``Knock'' sounds. Therefore, we submitted a fourth system, which is an ensemble of the other three.
This ensemble strategy follows the same approach as the temporal ensemble. However, for the ``Water tap'', ``Bell'', and ``Knock'' classes, if any one of the three models detects the sound, we consider it active, even if the condition of having at least 2 out of 3 predictions is not met.

\subsection{Results}
\label{sec:results}

The results achieved on the development set of the STARSS23 \cite{Shimada2023STARSS23AA} dataset are reported in Table\,\ref{tab:results}.
The AV-Conformer method achieves the highest F1 score, while the Depth-cued model has the lowest RDE. However, the Depth-cued model's F1 score is approximately 10 percentage points lower than that of the AV-Conformer, mainly due to its failure to detect certain rare classes, like ``Knock''.
The ensemble system recovers these 10 points in the F1 score. However, it does not provide the improvement in localization accuracy that is achieved with the temporal ensemble.
All submitted systems achieve superior performance than the audio-only (AO) and audio-visual (AV) baselines that employ the FOA audio format.

\begin{table}[tb]
\caption{Results on the development set of STARSS23 \cite{Shimada2023STARSS23AA}.}
\vspace{-2mm}
\begin{center}
%\footnotesize
\begin{tabular}{c|c|c|c} \hline

\textbf{Method}&$F_{\leq 20^{\circ}/1}\!\uparrow$& DOAE $\downarrow$ & RDE $\downarrow$ \\ 
\hline

Baseline AO & 13.1\% & 36.9° & 33.0\%  \\
Baseline AV & 11.3\% & 38.4° & 36.0\% \\
\hline
AV-Conformer & \textbf{40.8\%} & 17.7° & 30.5\% \\
AV-Conformer (TE) & 38.7\% & \textbf{16.8°} & 30.4\% \\
Depth-cued & 30.7\% & 18.9° & \textbf{27.0\%} \\
Ensemble & 40.3\% & 18.0° & 29.7\% \\
\hline
\end{tabular}
\label{tab:results}
\end{center}
\vspace{-7mm}
\end{table}

\section{Conclusion}
\label{sec:conclusion}

This technical report describes the 4 systems submitted to the Task 3 of DCASE 2024 Challenge (Track B). Specifically, two main models are explored: an AV-Conformer and a Depth-cued model, which correspond respectively to system one and system three of the four submitted.
The second system applies a temporal ensemble strategy to the output of the AV-Conformer, and the fourth system consists of an ensemble of the other three.
All our systems outperform the audio-only and audio-visual challenge baselines on the development set of the STARSS23 dataset.

\vspace{-3mm}
\section{ACKNOWLEDGMENT}
\label{sec:ack}

This research was funded by EPSRC-BBC Prosperity Partnership `{AI4ME}: Future personalised object-based media experiences delivered at scale anywhere' (EP/V038087/1). 
For the purpose of open access, the authors have applied a Creative Commons Attribution (CC BY) license to any Author Accepted Manuscript version arising. 
Data supporting this study are available from \url{https://zenodo.org/records/7880637}

% -------------------------------------------------------------------------
% Either list references using the bibliography style file IEEEtran.bst
\bibliographystyle{IEEEtran}
\bibliography{refs}
%
% or list them by yourself
% \begin{thebibliography}{9}
% 
% \bibitem{dcase2016web}
%   \url{http://www.cs.tut.fi/sgn/arg/dcase2016/}.
%
% \bibitem{IEEEPDFSpec}
%   {PDF} specification for {IEEE} {X}plore$^{\textregistered}$,
%   \url{http://www.ieee.org/portal/cms_docs/pubs/confstandards/pdfs/IEEE-PDF-SpecV401.pdf}.
%
% \bibitem{PDFOpenSourceTools}
%   Creating high resolution {PDF} files for book production with 
%   open source tools, 
%   \url{http://www.grassbook.org/neteler/highres_pdf.html}.
%
% \bibitem{eWilliams1999}
% E. Williams, \emph{Fourier Acoustics: Sound Radiation and Nearfield Acoustic
%   Holography}. London, UK: Academic Press, 1999.
% 
% \bibitem{ieeecopyright}
%   \url{http://www.ieee.org/web/publications/rights/copyrightmain.html}.
%
% \bibitem{cJones2003}
% C. Jones, A. Smith, and E. Roberts, ``A sample paper in conference
%   proceedings,'' in \emph{Proc. IEEE ICASSP}, vol. II, 2003, pp. 803--806.
% 
% \bibitem{aSmith2000}
% A. Smith, C. Jones, and E. Roberts, ``A sample paper in journals,'' 
%   \emph{IEEE Trans. Signal Process.}, vol. 62, pp. 291--294, Jan. 2000.
% 
% \end{thebibliography}

\end{sloppy}
\end{document}